\documentclass[reprint,prl,twocolumn,superscriptaddress,showpacs,showkeys,floatfix,preprintnumbers]{revtex4-1}

\pdfoutput=1

\usepackage{braket}
\usepackage{xargs}
\usepackage{mathtools}
\usepackage[english]{babel}
\usepackage[utf8]{inputenc}
\usepackage{amsmath,amssymb,braket,slashed,mathrsfs}
\usepackage{pifont}
\usepackage{MnSymbol}
\usepackage{graphicx}
\usepackage{tikz}
\usetikzlibrary{shapes,arrows,positioning}
\tikzset{decision/.style={diamond, draw, fill=blue!20, text width=4.5em, text badly centered, inner sep=0pt}}
\tikzset{block/.style={rectangle, draw, fill=blue!20, text width=10em, text centered, rounded corners, minimum width=3.5cm}}
\tikzset{block1/.style={rectangle, draw, fill=blue!20, text width=18.5em, text centered, rounded corners, minimum width=3.5cm}}
\tikzset{line/.style={draw, -latex, thick}}
\usepackage{color,hyperref}
\usepackage{multirow,array}

\usepackage{cleveref}
\newcommand{\ba}{\begin{eqnarray}}
\newcommand{\ea}{\end{eqnarray}}
\newcommand{\be}{\begin{equation}}
\newcommand{\ee}{\end{equation}}
\newcommand{\bi}{\begin{itemize}}
\newcommand{\ei}{\end{itemize}}

\newcommand{\nn}{\nonumber}

\newcommand{\innovation}{Collaborative Innovation Center of Quantum Matter, Beijing 100871, China}
\newcommand{\chep}{Center for High Energy Physics, Peking University, Beijing 100871, China}
\newcommand{\pkuphy}{School of Physics, Peking University, Beijing 100871,
China}
\newcommand{\KeyLab}{State Key Laboratory of Nuclear Physics and Technology,
Peking University, Beijing 100871, China}
\newcommand{\Uconn}{Department of Physics, University of Connecticut, Storrs, CT 06269, USA}
\newcommand{\RBRC}{RIKEN-BNL Research Center, Brookhaven National Laboratory, Building 510, Upton, NY 11973}
\newcommand{\Mainz}{Helmholtz Institute Mainz, Mainz, Germany}
\newcommand{\GSI}{GSI Helmholtzzentrum f\"ur Schwerionenforschung, Darmstadt, Germany}
\newcommand{\JGU}{Johannes Gutenberg University, Mainz, Germany}
\newcommand{\Bonn}{Helmholtz-Institut f\"ur Strahlen- und Kernphysik and Bethe
Center for Theoretical Physics,\\ Universit\"at Bonn, 53115 Bonn, Germany}

\begin{document}
\title{First-principles calculation of electroweak box diagrams from lattice QCD}

\author{Xu~Feng}\affiliation{\pkuphy}\affiliation{\innovation}\affiliation{\chep}\affiliation{\KeyLab}
\author{Mikhail~Gorchtein}\affiliation{\Mainz}\affiliation{\GSI}\affiliation{\JGU}
\author{Lu-Chang~Jin}\affiliation{\Uconn}\affiliation{\RBRC}
\author{Peng-Xiang~Ma}\affiliation{\pkuphy}
\author{Chien-Yeah~Seng}\affiliation{\Bonn}
%
\preprint{MITP/20-018}

\date{\today}

\begin{abstract}
We present the first realistic lattice QCD calculation of the $\gamma
W$-box diagrams relevant for beta decays. 
The nonperturbative low-momentum integral of the $\gamma W$ loop
is calculated using a lattice QCD simulation, 
complemented by the perturbative QCD result at high momenta.
Using the pion semileptonic decay as an example, we demonstrate the feasibility 
of the method.
By using domain wall fermions at the physical pion mass with
    multiple lattice spacings and volumes, we obtain the axial $\gamma W$-box
    correction to the semileptonic pion decay, $\Box_{\gamma
  W}^{VA}\big|_{\pi}=2.830(11)_{\mathrm{stat}}(26)_{\mathrm{sys}}\times10^{-3}$,
with the total uncertainty controlled at the level of $\sim1$\%. This
    study sheds light on the first-principles computation of the $\gamma W$-box correction
    to the neutron decay, which plays a decisive role in the determination of
    $|V_{ud}|$.
\end{abstract}

\maketitle
{\bf Introduction} --
The precise determination of the Cabibbo-Kobayashi-Maskawa (CKM) matrix elements, which
are fundamental parameters of the Standard Model, is
one of the central themes in modern particle physics.
In the CKM matrix, $V_{ud}$ is the most accurately-determined element
from the study of superallowed $0^+\to0^+$ nuclear beta decays~\cite{Hardy:2014qxa}
which are pure vector transitions at tree level and are theoretically clean due to
the protection of the conserved vector current. Going beyond tree level, the
electroweak radiative corrections
involving the axial-vector current become important and
ultimately dominate the theoretical uncertainties. 

Among various electroweak radiative corrections, the axial $\gamma W$-boson
box contribution $\Box_{\gamma W}^{VA}$ 
contains a significant sensitivity to low-energy
hadronic effects, and is a dominant source of the total theoretical
uncertainty~\cite{Sirlin:1977sv}.
The recent dispersive analysis~\cite{Seng:2018yzq,Seng:2018qru} 
reduced this uncertainty by a factor of 2 comparing to the previous study by Marciano and Sirlin~\cite{Marciano:2005ec}, and the updated result of 
$|V_{ud}|$ raised a 4 standard-deviation tension with the first-row CKM unitarity (barring possibly underestimated nuclear effects: see Ref.~\cite{Seng:2018qru,Gorchtein:2018fxl}).
The main difference between those works is the use of inclusive neutrino and antineutrino scattering data that Refs.
\cite{Seng:2018yzq,Seng:2018qru} used to estimate the contribution of 
the intermediate momenta inside the $\gamma W$ loop integral, $0.1\,\mbox{GeV}^2\lesssim Q^2\lesssim 1\,\mbox{GeV}^2$, 
prone to nonperturbative hadronic effects. 
To further improve the determination of $|V_{ud}|$, 
it requires either better-quality experimental input or the direct, precise
lattice QCD calculations of the $\gamma W$-box contribution. 

Lattice QCD has played an important role in the determination of the nonperturbative  
hadronic matrix elements needed to constrain the CKM unitarity. Recent lattice
results are averaged and summarized by the FLAG report 2019~\cite{Aoki:2019cca}. 
With lattice QCD simulations having reached an impressive level of
precision for tree-level parameters of the electroweak interaction, it becomes timely and important to study higher-order electroweak
corrections. The examples of such lattice applications include the QED corrections
to hadron
masses~\cite{Duncan:1996xy,Duncan:1996be,Blum:2007cy,Blum:2010ym,Ishikawa:2012ix,Borsanyi:2014jba,Boyle:2017gzv,Feng:2018qpx} 
and leptonic decay rates~\cite{Carrasco:2015xwa,Lubicz:2016xro,Giusti:2017dwk,DiCarlo:2019thl} and a series of
higher-order electroweak effects, such as $K_L$-$K_S$ mass
difference~\cite{Christ:2012se,Bai:2014cva,Wang:2020jpi},
$\epsilon_K$~\cite{Bai:2016gzv}, rare kaon
decays~\cite{Christ:2015aha,Christ:2016eae,Christ:2016mmq,Bai:2017fkh,Bai:2018hqu,Christ:2019dxu}
and double beta decays~\cite{Shanahan:2017bgi,Tiburzi:2017iux,Nicholson:2018mwc,Feng:2018pdq,Detmold:2018zan,Tuo:2019bue}. 
As for the $\gamma W$-box contribution, which is a QED correction to
semileptonic decays, it still remains a new horizon for lattice QCD.

It has been proposed to use the Feynman-Hellmann
theorem to calculate the $\gamma W$-box contribution~\cite{Bouchard:2016heu,Seng:2019plg}. 
In this work, we opt for a more straightforward way
to perform the lattice calculation. To demonstrate the feasibility of the
method, we carry out the exploratory study for the case of the pion
semileptonic decays. The calculation is performed at the physical pion mass with
various lattice spacings and volumes, which allows us to control the systematic
effects in the lattice results. Combining the results from lattice QCD together
with the 
perturbative QCD, we obtain the axial $\gamma W$-box correction
to pion decay amplitude with a relative $\sim1$\% uncertainty.

\begin{figure}[htb]
     \centering
         \includegraphics[width=0.35\textwidth,angle=0]{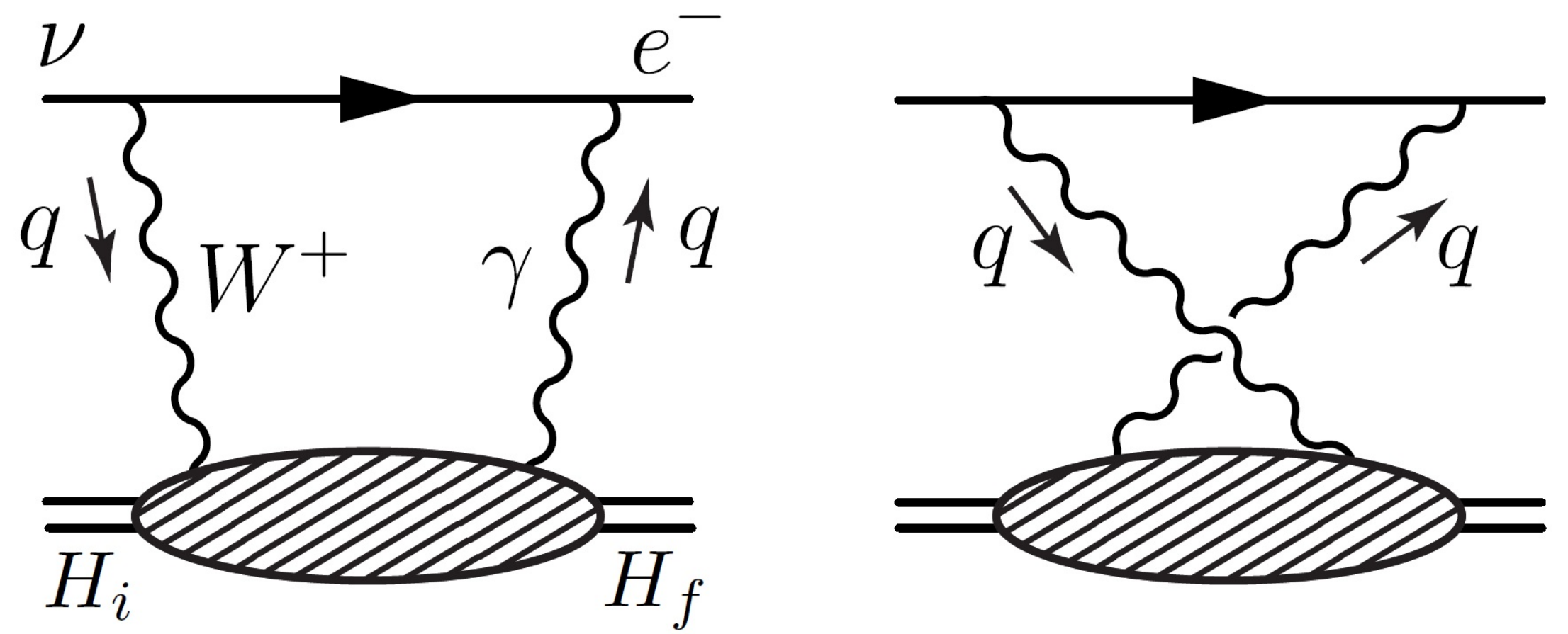}
         \caption{The $\gamma W$-box
         diagrams for the semileptonic decay process $H_i\to H_fe\bar{\nu}_e$.}
    \label{fig:photon_W_diags}
     \end{figure}

{\bf\boldmath The $\gamma W$-box contribution} -- In the theoretical analysis of the 
superallowed nuclear beta decay rates, the dominant uncertainty arises from the nucleus-independent
electroweak radiative correction, $\Delta_R^V$, which is universal for both nuclear and 
free neutron beta decay~\cite{Hardy:2014qxa}. Among various contributions to $\Delta_R^V$, Sirlin
established~\cite{Sirlin:1977sv} 
that only the axial $\gamma W$-box contribution is sensitive to hadronic scales;
see Fig.~\ref{fig:photon_W_diags} for the $\gamma W$ diagrams.
The relevant hadronic tensor $T^{VA}_{\mu\nu}$ is defined as
\be
T^{VA}_{\mu\nu}=\frac{1}{2}\int
d^4x\,e^{iqx}\langle
H_f(p)|T\left[J^{em}_\mu(x)J^{W,A}_\nu(0)\right]|H_i(p)\rangle,
\ee
for a semileptonic decay process $H_i\to
H_fe\bar{\nu}_e$. Above, $H_{i/f}$ are given by neutron and
proton for the neutron beta decay, and by $\pi^-$ and $\pi^0$ for the pion semileptonic
decay, respectively. Furthermore, 
$J^{em}_\mu=\frac{2}{3}\bar u \gamma_\mu u - \frac{1}{3} \bar d \gamma_\mu d-
\frac{1}{3} \bar s \gamma_\mu s$
is the electromagnetic quark current,
and $J^{W,A}_\nu=\bar u \gamma_\nu \gamma_5 d$ is the axial part
of the weak charged current.

The spin-independent part of  $T^{VA}_{\mu\nu}$ has only one term, 
$T^{VA}_{\mu\nu}=i\epsilon_{\mu\nu\alpha\beta}q^\alpha
p^\beta T_3+\dots$, where $T_3$ is a scalar function.
For the neutron beta decay, the spin-dependent contributions, denoted by the ellipses here, are absorbed into the definition of the
nucleon axial charge $g_A$, which can be measured directly from experiments.
According to current algebra~\cite{Sirlin:1977sv}, it is this spin-independent
term that gives rise to the hadron structure-dependent
contribution and dominates the uncertainty in the theoretical prediction.
Using $T_3$ as input,
the axial $\gamma W$-box correction to the tree-level amplitude is given as~\cite{Seng:2018yzq}
\ba
\label{eq:square}
&&\Box_{\gamma W}^{VA}\big|_H=\frac{1}{F_+^H}\frac{\alpha_e}{\pi}\int_{0}^{\infty}
dQ^2\,\frac{m_W^2}{m_W^2+Q^2}
\nn\\
&&\hspace{1.5cm}\times\int_{-\sqrt{Q^2}}^{\sqrt{Q^2}}\frac{dQ_0}{\pi}\frac{(Q^2-Q_0^2)^\frac{3}{2}}{(Q^2)^2}T_3(Q_0,Q^2).
\ea
Here $Q^2=-q^2>0$ is the spacelike four-momentum square. The normalization factor $F_+^H$
arises from the local matrix element $\langle
H_f(p')|J_\mu^{W,V}|H_i(p)\rangle=(p+p')_\mu F_+^H$, with $F_+^H=1$ for the neutron
and $\sqrt{2}$ for the pion decay.

{\bf Methodology} -- 
In the framework of lattice QCD, the hadronic tensor $T^{VA}_{\mu\nu}$ in Euclidean spacetime is given by
\be
T^{VA}_{\mu\nu}=\frac{1}{2}\int dt\, e^{-iQ_0t}\int
d^3x\,e^{-i\vec{Q}\cdot\vec{x}}\,\mathcal{H}^{VA}_{\mu\nu}(t,\vec{x})
\ee
with $\mathcal{H}^{VA}_{\mu\nu}(t,\vec{x})$ defined as
\be
\mathcal{H}^{VA}_{\mu\nu}(t,\vec{x})\equiv
\langle H_f(P)|T\left[J^{em}_\mu(t,\vec{x})J^{W,A}_\nu(0)\right]|H_i(P)\rangle.
\ee
Here the Euclidean momenta $P$ and $Q$ are chosen as
\be
P=(im_H,\vec{0}),\quad Q=(Q_0,\vec{Q})
\ee
with $m_H$ the hadron mass.

By multiplying $\epsilon_{\mu\nu\alpha\beta}Q_\alpha P_\beta$ to
$T^{VA}_{\mu\nu}$, we can extract the function $T_3(Q_0,Q^2)$ through
\be
\label{eq:T3}
T_3(Q_0,Q^2)=-\frac{\mathcal{I}}{2m_H^2|\vec{Q}|^2},
\quad
\mathcal{I}=\epsilon_{\mu\nu\alpha\beta}Q_\alpha
P_\beta
T^{VA}_{\mu\nu}.
\ee
Here $\mathcal{I}$ can be written in terms of
$\mathcal{H}^{VA}_{\mu\nu}$ as
\ba
\mathcal{I}&=&\frac{i}{2}\epsilon_{\mu\nu\alpha0}Q_\alpha m_H\int dt\,
e^{-iQ_0t}\int d^3\vec{x}\,e^{-i\vec{Q}\cdot\vec{x}}
\mathcal{H}^{VA}_{\mu\nu}
\nn\\
&=&\frac{m_H}{2}\int dt\,
e^{-iQ_0t}\int d^3\vec{x}\,e^{-i\vec{Q}\cdot\vec{x}}
\epsilon_{\mu\nu\alpha0}\frac{\partial \mathcal{H}^{VA}_{\mu\nu}}{\partial
x_\alpha}.
\ea
We can average over the spatial directions for $\vec{Q}$ and have
\ba
\label{eq:I}
\mathcal{I}&=&\frac{m_H}{2}\int dt\,e^{-iQ_0t}\int
d^3\vec{x}\,j_0\left(|\vec{Q}||\vec{x}|\right)\epsilon_{\mu\nu\alpha0}\frac{\partial \mathcal{H}^{VA}_{\mu\nu}}{\partial x_\alpha}
\nn\\
&=&\frac{m_H}{2}\int dt\,e^{-iQ_0t}\int
d^3\vec{x}\,\frac{|\vec{Q}|}{|\vec{x}|}j_1\left(|\vec{Q}||\vec{x}|\right)\epsilon_{\mu\nu\alpha0}x_\alpha\mathcal{H}^{VA}_{\mu\nu},
\nn\\
\ea
where $j_n(x)$ are the spherical Bessel functions.
A key ingredient in this approach is that once the Lorentz scalar function
$\epsilon_{\mu\nu\alpha0}x_\alpha\mathcal{H}^{VA}_{\mu\nu}$ is prepared, e.g. from
a lattice QCD calculation, one can determine $T_3(Q_0,Q^2)$ directly.

Putting Eqs.~(\ref{eq:I}) and (\ref{eq:T3}) into Eq.~(\ref{eq:square}) and changing the variables
as $|\vec{Q}|=\sqrt{Q^2}\cos\theta$ and $Q_0=\sqrt{Q^2}\sin\theta$, we obtain the master
formula 
\be
\label{eq:master}
\Box_{\gamma
W}^{VA}\big|_H=\frac{3\alpha_e}{2\pi}\int\frac{dQ^2}{Q^2}\,\frac{m_W^2}{m_W^2+Q^2}M_H(Q^2)
\ee
with
\ba
\label{eq:lattice_M}
&&M_H(Q^2)=-\frac{1}{6}\frac{1}{F_+^H}\frac{\sqrt{Q^2}}{m_H}\int
d^4x\,\omega(t,\vec{x})\epsilon_{\mu\nu\alpha0}x_\alpha\mathcal{H}^{VA}_{\mu\nu}(t,\vec{x}),
\nn\\
&&\omega(t,\vec{x})=\int_{-\frac{\pi}{2}}^{\frac{\pi}{2}}\frac{\cos^3\theta\,d\theta}{\pi}\frac{j_1\left(\sqrt{Q^2}|\vec{x}|\cos\theta\right)}{|\vec{x}|}
\cos\left(\sqrt{Q^2}t\sin\theta\right).
\nn\\
\ea
For small $Q^2$, lattice QCD can determine the
function $M_H(Q^2)$ with lattice discretization errors under control.

For large $Q^2$, we utilize the operator product expansion
\ba
\label{eq:OPE}
&&\frac{1}{2}\int d^4x e^{-iQx}T\left[J_{\mu}^{em}(x)J_{\nu}^{W,A}(0)\right]
\nn\\
&=&\frac{i}{2Q^2}\left\{C_a(Q^2)\delta_{\mu\nu}Q_\alpha-C_b(Q^2)\delta_{\mu\alpha}
Q_\nu\right.
\nn\\
&&\hspace{2cm}\left.-C_c(Q^2)\delta_{\nu\alpha} Q_\mu\right\}J^{W,A}_\alpha(0)
\nn\\
&&+\frac{1}{6Q^2}C_d(Q^2)\epsilon_{\mu\nu\alpha\beta}Q_\alpha
J^{W,V}_\beta(0)+\cdots.
\ea
There are only four possible local operators at leading twist.
(For the pion decay, the hadronic matrix elements for the first three operators
vanish.) Multiplying $\epsilon_{\mu\nu\alpha\beta}Q_\alpha P_\beta$ to the
relation~(\ref{eq:OPE}) we obtain
\ba
\label{eq:PT}
&&T_3(Q_0,Q^2)=\frac{C_d(Q^2)}{3Q^2}F_+^H+\cdots,
\nn\\
&&M_H(Q^2)=\frac{C_d(Q^2)}{12}+\cdots,
\ea
where the ellipses remind us that the higher-twist contributions are not included yet. The
Wilson coefficient $C_d(Q^2)$ is calculated to four-loop
accuracy~\cite{Larin:1991tj,Baikov:2010je}
\be
C_d(Q^2)=\sum_n c_n a_s^n,\quad a_s=\frac{\alpha_s(Q^2)}{\pi},
\ee
with coefficients $c_n$ given in Eq.~(12) of Ref.~\cite{Baikov:2010je}.
Here $\alpha_s$ is the strong coupling constant.

We introduce a momentum-squared scale $Q^2_{\mathrm{cut}}$ that separates the two regimes, 
and
split the integral in Eq.~(\ref{eq:master}) into two parts
\ba
\Box_{\gamma W}^{VA}\big|_H&=&\Box_{\gamma
W}^{VA,\le}\big|_H+\Box_{\gamma W}^{VA,>}\big|_H
\nn\\
&=&\left(\int_0^{Q^2_\mathrm{cut}}\frac{dQ^2}{Q^2}+\int_{Q^2_\mathrm{cut}}^\infty
\frac{dQ^2}{Q^2}\right)\frac{m_W^2}{m_W^2+Q^2}M_H(Q^2).
\nn\\
\ea
With Eq.~(\ref{eq:lattice_M}) we use the lattice data to determine the integral for $Q^2\le
Q^2_{\mathrm{cut}}$, while with Eq.~(\ref{eq:PT}) we use perturbation theory to determine the integral for
$Q^2>Q^2_{\mathrm{cut}}$.

     \begin{figure}[htb]
     \centering
         \includegraphics[width=0.48\textwidth,angle=0]{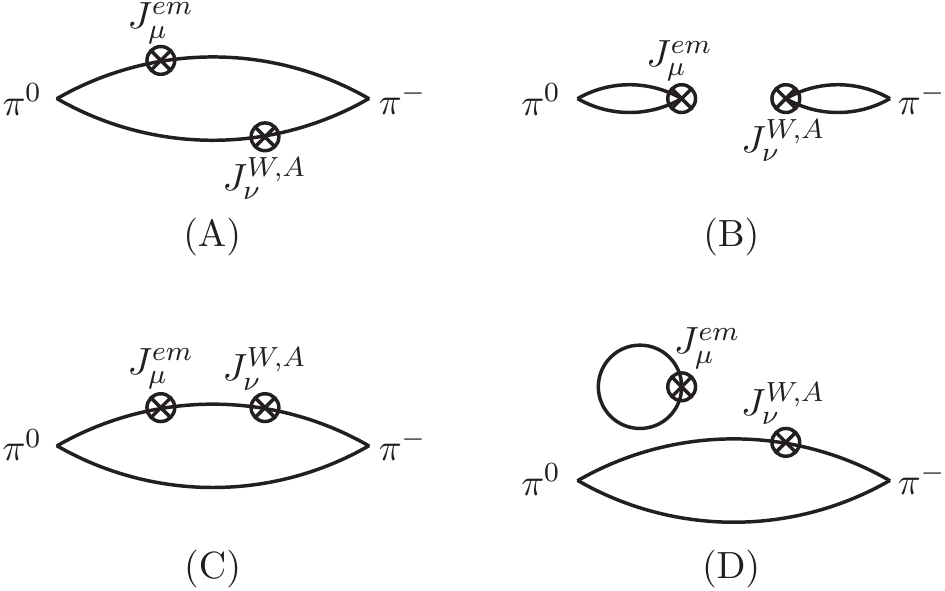}
         \caption{Four types of quark contractions for pion $\gamma W$-box
         diagrams.}
    \label{fig:contraction}
     \end{figure}

{\bf Lattice setup} -- 
We use five lattice QCD gauge ensembles at the physical pion mass,
generated by RBC and UKQCD Collaborations using $2+1$-flavor domain wall fermion~\cite{Blum:2014tka}. The ensemble
parameters are shown in Table~\ref{tab:ensemble_parameter}.
Here 48I and 64I
use the Iwasaki gauge action in the simulation (denoted as Iwasaki in this work) while the other three ensembles use
Iwasaki+DSDR action (denoted as DSDR).
We calculate the correlation function
$\langle \phi_{\pi^0}(t_f)J_\mu^{em}(x)J_\nu^{W,A}(y)\phi_{\pi^-}^\dagger(t_i)\rangle$ with
$t_f=\operatorname{max}\{t_x,t_y\}+\Delta t$ and
$t_i=\operatorname{min}\{t_x,t_y\}-\Delta t$.
We use the wall-source pion
interpolating operators $\phi_{\pi^0}$ and $\phi_{\pi^-}^\dagger$, which
have a good overlap with the $\pi$ ground state, and find the ground-state
saturation for $\Delta t\gtrsim1$ fm. In practice the
values of $\Delta t$ are chosen conservatively as shown in Table~\ref{tab:ensemble_parameter}.
For each ensemble we use the gauge configurations with sufficiently
long separation, i.e. each separated by at least
10 trajectories. The number
of configurations used is listed in Table~\ref{tab:ensemble_parameter}.

\begin{table}[htbp]
	\small
	\centering
	\begin{tabular}{cccccccc}
		\hline
        \hline
        Ensemble  & $m_\pi$ [MeV] & $L$ &  $T$ & $a^{-1}$ [GeV]&
        $N_{\text{conf}}$ & $N_r$  & $\Delta t/a$ \\
        \hline
        24D  & 141.2(4) & $24$ & $64$ & $1.015$ & 46 & 1024  & 8  \\
        32D  & 141.4(3) & $32$ & $64$ & $1.015$ & 32  & 2048 & 8 \\
        32D-fine & 143.0(3) & $32$ & $64$ & $1.378$ & 71 & 1024  & 10 \\
        48I & 135.5(4) & $48$ & $96$ & $1.730$ & 28 & 1024 & 12 \\
        64I & 135.3(2) & $64$ & $128$ & $2.359$ & 62 & 1024  & 18 \\
        \hline
    \end{tabular}%
    \caption{Ensembles used in this work. For each ensemble we list the pion mass $m_\pi$, 
    the spatial and temporal extents, $L$ and $T$, 
    the inverse of lattice
    spacing $a^{-1}$, the number
    of configurations used, $N_{\text{conf}}$, the number of point-source
    light-quark propagator
    generated for each configuration, $N_r$, and the
    time separation, $\Delta t$, used for the $\pi$
    ground-state saturation.}
    \label{tab:ensemble_parameter}%
\end{table}%

There are four types of contractions for $\gamma W$-box diagrams as shown in
Fig.~\ref{fig:contraction}. 
We produce wall-source quark propagators on all time slices. 
Using the techniques described in Ref.~\cite{Tuo:2019bue} 
type (A) and (B) diagrams can be calculated with high
precision by performing the spacetime-translation average over $L^3\times T$
measurements.
Under the $\gamma_5$ Hermitian conjugation of
the Euclidean quark propagators, one can confirm that type (B) does not
contribute to the axial $\gamma W$-box diagrams. Type (C) diagram is calculated
by treating one current as the source and the other as the sink. 
We calculate point-source propagators at $N_r$ random spacetime locations. 
The values of $N_r$ are shown in Table~\ref{tab:ensemble_parameter}.
These point-source propagators can be placed at either electromagnetic
current or weak current. We 
thus average the type (C) correlation functions over $2\,N_r$ measurements. This
is similar with the treatment taken by Ref.~\cite{Feng:2019geu}.
We neglect the disconnected contribution (D), which vanishes in the flavor
$SU(3)$ limit.

   \begin{figure}[htb]
     \centering
         \includegraphics[width=0.48\textwidth,angle=0]{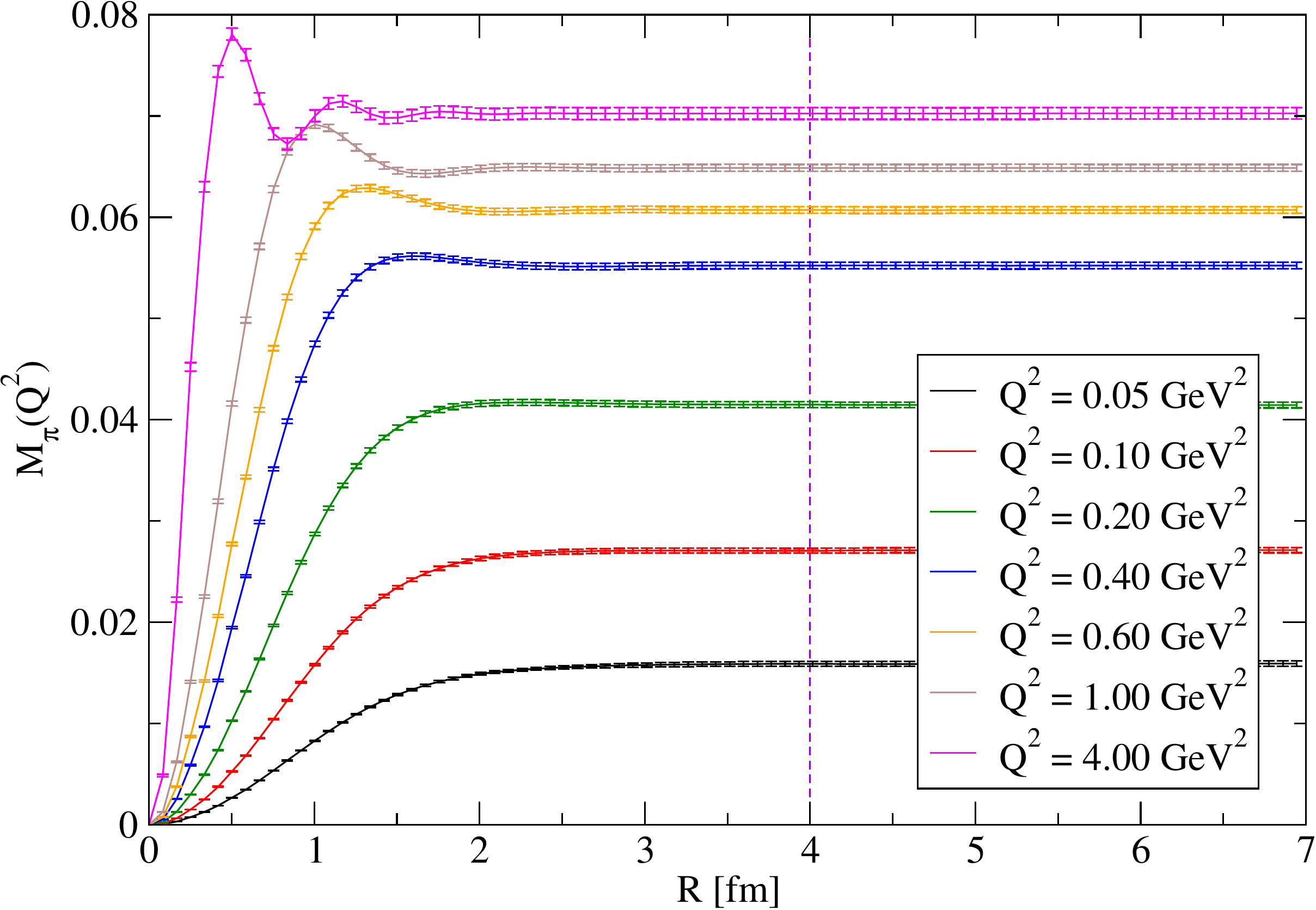}
         \caption{For ensemble 64I, lattice results of $M_\pi(Q^2)$ as a function of
         the integral range $R$.}
         \label{fig:R_dep}
    \end{figure}

{\bf Numerical results} --
In practice, the integral in Eq.~(\ref{eq:lattice_M}) can be performed within a range of
$\sqrt{t^2+\vec{x}^2}\le R$. 
Taking the ensemble 64I as an example, $M_\pi(Q^2)$ as a function of the integral range $R$ is shown in
Fig.~\ref{fig:R_dep}. We find that for all the momenta $Q^2\in[0,4]$ GeV$^2$, the
integral is saturated at large $R$. 
We choose the truncation range $R_0\simeq 4$ fm, 
which is a conservative choice for all ensembles listed in Table~\ref{tab:ensemble_parameter}.
The contributions to the integral from $\sqrt{t^2+\vec{x}^2}> R_0$ is negligible,
indicating that the finite-volume effects are well under control in our
calculation. We can further verify this conclusion by a direct comparison using the 24D
and 32D data.

   \begin{figure}[htb]
     \centering
         \includegraphics[width=0.48\textwidth,angle=0]{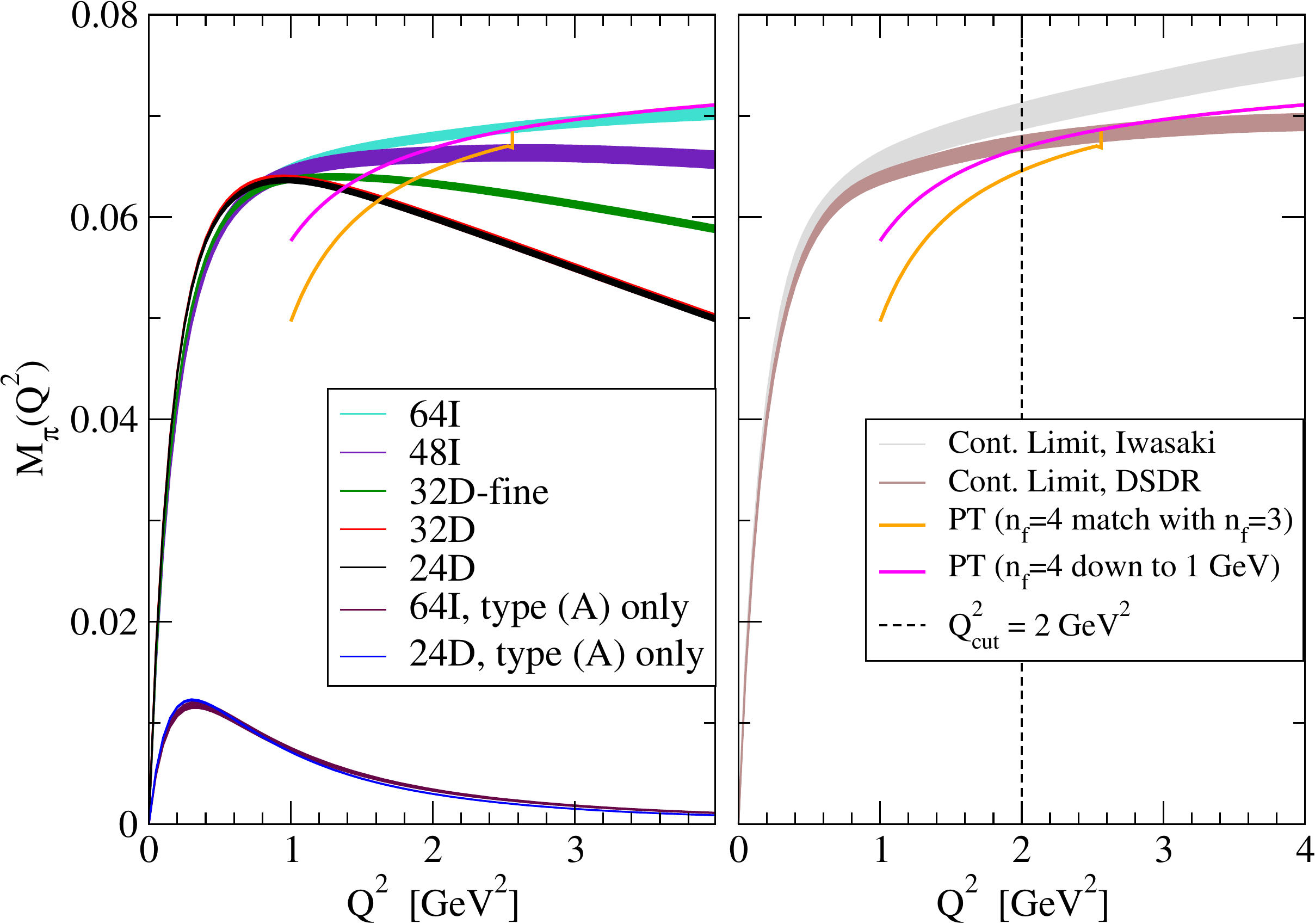}
         \caption{$M_\pi(Q^2)$ as a function of $Q^2$. In the left panel, 
         the lattice results for ensembles 64I, 48I, 32D-fine, 32D and 24D
         are represented by
         turquoise, indigo, dark green, red and black bands, respectively. 
         Taking 64I and 24D as examples, the results for type (A)
         diagram are also plotted. In the right panel, Iwasaki and DSDR results at the continuum limit are
         shown by the gray and brown bands.
         The orange curve shows the 
         results from perturbation theory by decoupling the charm quark at 1.6 GeV while the magenta one
         is compiled using the 4-flavor perturbation theory continuously down to 1 GeV.
         }
         \label{fig:m_qsq}
    \end{figure}

The lattice results of $M_\pi(Q^2)$ as a function of $Q^2$ are shown in
Fig.~\ref{fig:m_qsq} together with the perturbative results.
Ensemble 24D and 32D have the same pion mass and
lattice spacing but different volumes. The good agreement between these two
ensembles indicates that the finite-volume effects are smaller than the statistical errors. 
At $Q^2\gtrsim 1$ GeV$^2$, the lattice discretization effects dominate the
uncertainties. In the left panel of Fig.~\ref{fig:m_qsq} an obvious discrepancy
is observed at large $Q^2$ for the lattice results with
different lattice spacings.

For the perturbation theory, the Wilson coefficient $C_d(Q^2)$ is determined using
the RunDec package~\cite{Chetyrkin:2000yt}, where $\alpha_s$ is calculated to four-loop accuracy.
At low $Q^2$ the results still contain large systematic
uncertainties
due to the lack of higher-loop and higher-twist contributions.
In Fig.~\ref{fig:m_qsq} we show two curves from perturbation theory. One is
compiled using 4-flavor theory down to 1 GeV, while the other decouples the charm
quark at 1.6 GeV and uses 3-flavor theory for $(\mbox{1 GeV})^2\le
Q^2\le(\mbox{1.6 GeV})^2$. The discrepancy between the two curves suggests an
O(14\%) systematic effect in the perturbative determination of $M_H(Q^2)$ at $Q^2\approx1$ GeV$^2$.

{\bf Estimate of systematic effects} -- For $\Box_{\gamma W}^{VA,\le}$ the largest uncertainties arise from the
lattice discretization effects.
Since Iwasaki and DSDR ensembles have different lattice discretizations, we 
treat them separately. After the linear extrapolation in
$a^2$, the
Iwasaki and DSDR results at continuum limit are shown in the right panel of
Fig.~\ref{fig:m_qsq}. Using $Q_{\mathrm{cut}}^2=2$ GeV$^2$ we obtain
\be
\Box_{\gamma W}^{VA,\le}\big|_\pi=
\begin{cases}
    0.671(11)\times10^{-3} & \mbox{for Iwasaki} \\
    0.647(7)\times10^{-3} & \mbox{for DSDR}
\end{cases}.
\ee
We take the Iwasaki result as the central value
and estimate the residual $O(a^4)$ lattice artifacts using the discrepancy between Iwasaki
and DSDR.

For $\Box_{\gamma W}^{VA,>}$ the largest uncertainties 
arise from the higher-loop and higher-twist truncation effects.  
We estimate the former by comparing the 4-loop and 3-loop results from
perturbation theory. For the latter, unfortunately the complete information 
is not available. Considering the fact that type (A)
diagram, which has
two currents located at different quarks lines, only contains the higher-twist
contributions, we use it to estimate the size of higher twist. 
At $Q_{\mathrm{cut}}^2=2$ GeV$^2$ we have
\be
\Box_{\gamma
W}^{VA,>}\big|_\pi=2.159(6)_{\mathrm{HL}}(7)_{\mathrm{HT}}\times10^{-3},
\ee
where the central value is compiled using the 4-flavor theory (see the magenta curve
in the right panel of Fig.~\ref{fig:m_qsq}).
The first error indicates the higher-loop effects.
The second one stands for the higher-twist effects, which are compiled from the integral 
of $Q^2>Q_{\mathrm{cut}}^2$ using the type
(A) data as input.

{\bf Summary of results} -- After combining the results of $\Box_{\gamma
W}^{VA,\le}$ from lattice QCD and
$\Box_{\gamma W}^{VA,>}$ from perturbation theory, we obtain the total
contribution of $\Box_{\gamma W}^{VA}$
\ba
\label{eq:total}
\Box_{\gamma
W}^{VA}\big|_\pi&=&2.830(11)_{\mathrm{stat}}(9)_{\mathrm{PT}}(24)_{\mathrm{a}}(3)_{\mathrm{FV}}\times10^{-3}
\nn\\
&=&2.830(11)_{\mathrm{stat}}(26)_{\mathrm{sys}}\times10^{-3},
\ea
where the first uncertainty is statistical, and the remaining errors account for
perturbative truncation and higher-twist effects,
lattice discretization effects,
and lattice finite-volume effects by comparing the 24D and 32D results.
We add these systematic errors in quadrature to obtain the final systematic error.
For comparison, we also calculate $\Box_{\gamma
W}^{VA}\big|_\pi=2.816(9)_{\mathrm{stat}}(24)_{\mathrm{PT}}(18)_{\mathrm{a}}(3)_{\mathrm{FV}}\times10^{-3}$ at
$Q_{\mathrm{cut}}^2=1$ GeV$^2$ and $\Box_{\gamma
W}^{VA}\big|_\pi=2.835(12)_{\mathrm{stat}}(5)_{\mathrm{PT}}(30)_{\mathrm{a}}(3)_{\mathrm{FV}}\times10^{-3}$ at
$Q_{\mathrm{cut}}^2=3$ GeV$^2$. Both results are consistent with
Eq.~(\ref{eq:total}). 

For the pion semileptonic decay, the PIBETA experiment~\cite{Pocanic:2003pf} has improved the measurement of 
the branching ratio to 0.6\%.
The Standard Model prediction of the decay rate is given
by~\cite{Sirlin:1977sv,Pocanic:2003pf}
\be
\Gamma_{\pi\ell3}=\frac{G_F^2|V_{ud}|^2m_\pi^5|f_+^\pi(0)|^2}{64\pi^3}\left(1+\delta\right)I_\pi
\ee
with $G_F=1.1663787(6)\times 10^{-5}\:\mathrm{GeV}^{-2}$ the Fermi's  constant measured from the muon decay, $m_\pi$ the charged pion mass, $f_+^\pi(0)=1$ the
tree-level semileptonic form factor and $I_\pi=7.376(1)\times10^{-8}$ a known
kinematic factor. Numerically, $\Gamma_{\pi l3}=0.3988(23)\:\mathrm{s}^{-1}$
after taking into account the updated value of $\pi^+\rightarrow e^+\nu(\gamma)$
branching ratio as an overall normalization~\cite{Czarnecki:2019iwz}. The
effects of radiative corrections are contained in $\delta$. 
The existing analysis from chiral perturbation theory (ChPT) yields
$\delta=0.0334(10)_{\mathrm{LEC}}(3)_{\mathrm{HO}}$~\cite{Jaus:1999dw,Cirigliano:2002ng,Passera:2011ae,Czarnecki:2019iwz} 
with an overall theoretical uncertainty of
$\Gamma_{\pi\ell3}$ at a level of $0.1\%$. 
Here the first error is from the low energy constants and the second
is the uncertainty in determining the higher-order QED
effects~\cite{Erler:2002mv}.
Thus the experimental
measurement dominates the uncertainties and results in the determination of
$|V_{ud}|=0.9739(28)_\mathrm{exp}(5)_\mathrm{th}$ with a 0.3\% uncertainty.

We now show how our calculation reduces the uncertainty in $\delta$.
We adopt Sirlin's parameterization~\cite{Sirlin:1977sv} with slight modifications:
\be
\label{eq:delta}
\delta=\frac{\alpha_e}{2\pi}\left[\bar{g}+3\ln\frac{m_Z}{m_p}+\ln\frac{m_Z}{m_W}+\tilde{a}_g\right]+\delta_\mathrm{HO}^\mathrm{QED}+2\Box_{\gamma W}^{VA}.
\ee
By separating the axial $\gamma W$-box part into $\Box_{\gamma W}^{VA}$,
the remaining contributions are model independent and are given as follows.
\bi
\item Sirlin's function $\bar{g}$ arises from a
structure-indepenent, UV-finite one-loop integral. It accounts for
the infrared contributions involving the vector $\gamma
W$-box and the bremsstrahlung corrections. 
It contains a $3\ln m_p$ term that cancels the $m_p$-dependence in
$3\ln \frac{m_Z}{m_p}$.
Here $m_p$ is the proton mass that appears just as a matter of convention.
Numerically, one has
$\frac{\alpha_e}{2\pi}\bar{g}=1.051\times 10^{-2}$~\cite{Sirlin:1977sv,Wilkinson:1970cdv}. 
\item $\tilde{a}_g$ represents the $O(\alpha_s)$ QCD correction to all one-loop
    diagrams except for the axial
$\gamma W$ box. The integral in $\tilde{a}_g$ is
dominated by the high-energy scale $Q^2\simeq m_W^2$, where $\alpha_s$ is
small. As a consequence
$\frac{\alpha_e}{2\pi}\tilde{a}_g\approx -9.6\times10^{-5}$ is a small
contribution~\cite{Sirlin:1977sv,Seng:2019lxf}.
\item $\delta_{\mathrm{HO}}^\mathrm{QED}=0.0010(3)$ summarizes the leading-log
    higher-order QED effects which can be accounted for through the running of
    $\alpha_e$. The uncertainty assignment follows Ref.~\cite{Erler:2002mv}.
\ei    
Although the detailed uncertainties for $\bar{g}$ and $\tilde{a}_g$ are not given, by
power counting the intrinsic precision for the terms in the square brackets
(multiplied by $\frac{\alpha_e}{2\pi}$) is of the order
$G_Fm_p^2\sim10^{-5}$.

Combining the $\Box_{\gamma W}^{VA}$ in Eq.~\eqref{eq:total}, we now obtain
\begin{equation}
    \label{eq:correction}
    \delta = 0.0332(1)_{\gamma W}(3)_{\mathrm{HO}},
\end{equation} 
which corresponds to an almost complete removal of the dominant LEC 
uncertainties in the ChPT expression, and a reduction of the total 
uncertainty by a factor of 3.
Therefore, any theoretical improvement in the future will unavoidably
require a complete electroweak two-loop analysis. 
Consequently, the $|V_{ud}|$ determined from the pion semileptonic decay now reads: $|V_{ud}|=0.9740(28)_\mathrm{exp}(1)_\mathrm{th}$.

{\bf Conclusion} --
In this work we perform the first realistic lattice QCD calculation of the
$\gamma W$-box correction to the pion semileptonic decay, $\Box_{\gamma
W}^{VA}\big|_\pi$. 
The final result combines the lattice data at low momentum and perturbative
calculation at high momentum.
We use multiple
lattice spacings and volumes at the physical pion mass to control the continuum
and infinite-volume limits and obtain $\Box_{\gamma W}^{VA}\big|_\pi$ with
a total error of $\sim1$\%. As a result,
the uncertainty of the theoretical prediction for the pion semileptonic decay rates is reduced by a factor of $3$.
This result does not impact the first-row CKM unitarity due to the large 
experimental error, but a follow-up work~\cite{Seng:2020wjq} shows that the 
4-$\sigma$ tension persists.

The combined experimental measurement of 14 nuclear superallowed beta decays \cite{Hardy:2014qxa} is 10 times
more accurate than the current pion semileptonic decay experiment. On the other hand, 
the free neutron decay \cite{Tanabashi:2018oca,Markisch:2018ndu} leads to a 4.5 times better precision. 
In these two cases, the nonperturbative, structure-dependent $\gamma W$-box
contribution plays a decisive role. 
The technique presented in this work can be straightforwardly
generalized to a lattice calculation of the nucleon $\gamma W$-box corrections,
which are universal for both free and bound neutron decay. 
The latter is the key to a precise determination of
$|V_{ud}|$ and a stringent test of CKM unitarity.

\begin{acknowledgments}
{\bf Acknowledgments} -- X.F. and L.C.J. gratefully acknowledge many helpful discussions with our colleagues from the
RBC-UKQCD Collaborations.
We thank Yan-Qing Ma and Guido Martinelli for inspiring discussions.
X.F. and P.X.M. were supported in part by NSFC of China under Grant No. 11775002.
M.G.  is supported by EU Horizon 2020 research and innovation programme, 
STRONG-2020 project, under grant agreement No 824093 and  by the German-Mexican 
research collaboration Grant No. 278017 (CONACyT) and No. SP 778/4-1 (DFG).
L.C.J. acknowledges support by DOE grant DE-SC0010339. 
The work of C.Y.S. is supported in part by the DFG 
(Grant No. TRR110) and the NSFC (Grant No. 11621131001) through the funds 
provided to the Sino-German CRC 110 Symmetries and the Emergence of Structure in QCD, 
and also by the Alexander von Humboldt Foundation through the Humboldt Research Fellowship.
The computation is performed under the ALCC Program of
the US DOE on the Blue Gene/Q (BG/Q) Mira computer at the Argonne Leadership Class Facility,
a DOE Office of Science Facility supported under Contract DE-AC02-06CH11357.
Computations for this work were carried out in part on facilities of the USQCD
Collaboration, which are funded by the Office of Science of the U.S. Department
of Energy.
The calculation is also carried out on Tianhe 3 prototype at Chinese National Supercomputer Center in Tianjin.
\end{acknowledgments}

\bibliography{paper}

\end{document}